\documentclass[11pt,twoside]{article}
\usepackage{CAGN2016}
\usepackage{graphicx}

\usepackage[T1]{fontenc} % Computer Modern (CM) fonts

\usepackage{latexsym}
\usepackage{verbatim}

\begin{document}

\vskip 1.0cm
\markboth{Moll{\'a} et al.}{The evolution of  Oxygen radial gradient}
\pagestyle{myheadings}
%
%%%%  USE THE LINE THAT DESCRIBES THE CHARACTER OF YOUR WORK %%%%%%
%
\vspace*{0.5cm}
\parindent 0pt{Invited Review}
%\parindent 0pt{Contributed  Paper}
%\parindent 0pt{Poster}

%\vskip 0.3cm

\vspace*{0.5cm}
\title{The  evolution  of the radial gradient of Oxygen abundance in spiral galaxies}

\author{M.~Moll{\'a}$^{1}$, A.I.~D{\'\i}az$^{2,3}$, Y.~Ascasibar$^{2,3}$, B.K.~Gibson$^{4}$, O.~ Cavichia$^{5}$, R.D.D.~Costa$^{6}$, and W.J.~Maciel$^{6}$}
\affil{$^1$ CIEMAT, Avda. Complutense 40, 28040 Madrid, Spain\\
$^{2}$ Universidad Aut\'{o}noma de Madrid, 28049, Madrid, Spain \\
$^{3}$ Astro-UAM, Unidad Asociada CSIC, Universidad Aut{\'o}noma de Madrid,28049, Madrid, Spain
$^{4}$ E.A Milne Centre for Astrophysics, University of Hull, HU6~7RX, United Kingdom\\
$^5$ Instituto de F\'{i}sica e Qu\'{i}mica, Universidade Federal de Itajub\'{a}, Av. BPS, 1303, 37500-903, Itajub\'{a}-MG, Brazil\\
$^6$ Instituto de Astronomia, Geofisica e Ci{\^e}ncias Atmosf{\' e}ricas, Universidade de S{\^a}o Paulo, 05508-900, S{\^a}o Paulo-SP, Brazil}

\begin{abstract}
The aim of this work is to present our new series of chemical evolution models computed for spiral and low mass galaxies of different total masses and star formation efficiencies. We analyze the results of models, in particular the evolution of the radial gradient of oxygen abundance. Furthermore, we study the role of the infall rate and of the star formation history on the variations of this radial gradient.
The relations between the O/H radial gradient and other spiral galaxies characteristics as the size or the stellar mass are also
shown. We find that the radial gradient is mainly a scale effect which basically does not change with the redshift (or time) if it is measured within the optical radius. Moreover, when it is measured as a function of a normalized radius, show a similar value for all galaxies masses, showing a correlation with a dispersion around an average value
which is due to the differences star formation efficiencies, in agreement with the idea of a universal O/H radial gradient. 
\end{abstract}

\section{Introduction}
\label{intro}

The  elemental abundances  in  spiral and low mass galaxies are lower in the outer regions than in the inner ones, showing a well characterized radial gradient defined by the slope of a least-squares straight line fitted to the radial distribution of these abundances along the galactocentric radius (Shaver et al. 1983; Zaristky et al, 1994; Henry \& Worthey 1999). These radial gradients seem to correlate with other characteristics defining their galaxies. This way, they are flatter in the early galaxies than in the late ones. They also seem steeper in the low mass galaxies than in the bright massive disks. This radial gradient is considered as an evolutionary effect, that is, it comes from a difference of enrichment between regions more evolved (at the inner parts of disk) compared with the less evolves zones of the outer disks. This way, a flat gradient implies a more rapid evolution than in disks where the gradient is steeper, as shown in Moll{\'a}, Ferrini \& D{\'\i}az (1996) for a set of models for some nearby galaxies. These models resulted in a steep radial gradient for NGC~300 or M~33, while M~31 had a flatter gradient than our Milky Way Galaxy (MWG), and other similar galaxies as NGC~628 or  NGC~6946. Since the evolution modifies the level of enrichment of a given region or galaxy, it is expected that the radial gradient also changes with time and, therefore, that high-intermediate redshift galaxies would have to show a steeper radial gradient than in the present time, at least when measured as $dex\,kpc^{-1}$.  This results was obtained in Moll{\'a}, Ferrini \& D{\'\i}az (1997), further obtained later in Moll{\'a} \& D{\'\i}az (2005), hereafter MD05, and was also supported by the 
Planetary Nebulae (PN) O/H abundance data (Maciel, Costa \& Uchida 2003) and by open stellar cluster metallicities for different ages bins. Results by cosmological simulations for a MWG-like galaxy also obtained a similar behavior (Pilkington et al. 2012). 

However, when a correct feedback is included in these simulations, the radial gradient results to have a very similar slope for all times/redshifts (Gibson et al. 2013).
In turn, the most recent PN data (Stanghellini \& Haywood 2010, Maciel \& Costa 2013, Magrini et al. 2016) refined to estimate with more precision their ages and distances, give now  the same result: there is no evidences of  evolution of the radial gradient with time for MWG nor for other close spiral galaxies, at least until $z=1.5$. 
Simultaneously, there are, however, some recent observational data which estimate the abundances of galaxies at high and intermediate redshift, and which obtain a {\sl plethora} of different radial gradients with values as different as -0.30\,dex\,kpc$^{-1}$ or $+0.20$\,dex\,kpc$^{-1}$ (Cresci et al. 2010, Yuan et al. 2011, Queyrel et al. 2012, 
 Jones et al. 2013,  Genovali et al. 2014, Jones et al. 2015, Xiang et al. 2015, Anders et al. 2016). It is therefore necessary to revise our chemical evolution models and analyze in detail the evolution of this radial gradient not only for the MWG, but also for different galaxies.

\section{Chemical evolution model description}
\label{models}

We have computed a series of 76 models applied to spiral galaxies with dynamical masses in the range $5\times 10^{10}$--$10^{13}$\,$\rm M_{\odot}$ (with mass step
in logarithmic scale of  $\Delta\log{M}=0.03$) which implies disk total masses in the range $1.25\times10^{8}$--$5.3\times10^{11}$\,$\rm M_{\odot}$, or, equivalently rotation velocities between 42 and 320 km\,s$^{-1}$. The radial distributions of these masses are calculated through equations from Salucci et al.(2007), based in the rotation curves and their decomposition in halo and disk components. 

The scenario is the same as the one from MD05, with the total mass in a spherical regions at the time $t=0$, which infall over the equatorial plane and forms out the disk.
The gas infall rates are computed by taking into account the relationship between halo mass and disk mass in order to obtain at the end of the evolution disks as observed, with the adequate mass. They result to be higher in the centers of disks (bulges) and lower in disks, decreasing towards the outer regions, as expected 
in an inside-out scenario. However, the evolution with redshift is very similar for all disk regions, only with differences in the absolute values of infall rates, but
with a smooth decreasing for decreasing $z$, except for the central regions for which the infall changes strongly with $z$, more similarly to the cosmological
simulations results found for early and spheroidal galaxies. Details about these resulting rates are given in Moll{\'a} et al. (2016). 

Within each galaxy we assume that there is star formation (SF) in the halo, following a Schmidt-Kennicutt  power law on the gas density with an index $n=1.5$. In the disk, however, we have a star formation law in two steps: first molecular clouds form from diffuse gas, then stars form from cloud-cloud collisions (or by the interaction of
massive stars with the molecular clouds surrounding them). In our classical standard models from MD05, we treated these processes as depending on the volume of each region and a probability factor or efficiency for each one. For the halo SF, it is assumed an efficiency constant for all galaxies. The process of  interaction of massive stars with clouds is considered as local and we use the same approach for all galaxies. The two other efficiencies defining the molecular clouds and stars formation processes  are modified simultaneously from a model to another, with values between 0 and 1. 

In this new series of models we have also used an efficiency to form stars from molecular clouds, but to convert diffuse gas into molecular phase we have tried six different methods, two based in the same efficiency method as in MD05, called STD and MOD, and four based in different prescriptions based in Blitz \& Rosolowsky (2006), Krumhold et al (2008, 2009),  Gnedin \& Kravtsov (2011), and Ascasibar et al. (in preparation),  so-called BLI, KRU, GNE,  and ASC, respectively. More details about  these calculations and their implementation in our code are given in Moll{\'a} et al. (submitted), where we have applied the models to MWG and have checked which of these techniques give the best results when comparing the observational data.  Our results indicate that the technique ASC shows a behavior in better agreement with data than the others, In particular those related with the ratio $HI/H_{2}$ along radius or as a function of the gas density. 

The stellar yields are selected as derived in Moll{\'a} et al. (2015) among 144 different combinations with which we calculate a MWG model to see which of them 
is the best in reproducing the MWG data. We chose the Gavil{\'a}n et al (2005,2006) stellar yields for low and intermediate stars, the  ones from
Limongi \& Chieffi (2003) and Chieffi \& Limongi (2004) for massive stars, and  the Kroupa (2002) IMF. The Supernova type Ia yields from Iwamoto et al (1995) are also used. 

\section{Results}
\label{results}
\subsection{Evolution of the O/H radial gradient in MWG}

\begin{figure} 
\begin{center}
\includegraphics[width=9.0cm,angle=-90]{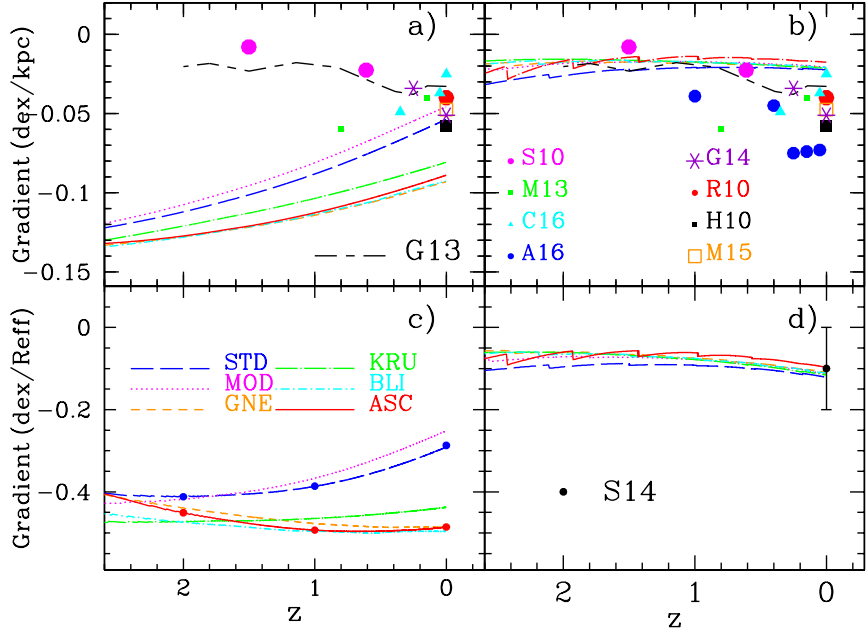}
\caption{Evolution of the O/H radial gradient with redshift $z$ for the MWG, measured as: top panels) dex\,kpc$^{-1}$ and bottom panels) as dex\,R$_{\rm eff}^{-1}$.  Solid lines represent our different models as labelled in c). Left panels show gradients for the whole radial range. Right panels show gradients only using regions within  $\rm R\le 2.5\,R_{\rm eff}$. Data are from   Stanghellini \& Haywood (2010, S10), Maciel \& Costa (2013, M13), Cunha et al. (2016, C16),  Anders et al. (2016, A16), Henry et al. (2010, H10), Rupke et al. (2010, R10), Genovali et al. (2014, G14), S{\'a}nchez et al.(2014, S14) and  Moll{\'a} et al. (2015, M15), as labelled in b) and d). Cosmological
simulations result for a MWG-like object from Gibson et al. (2013) is also drawn in a) and b). }
\label{mwg}
\end{center}
\end{figure}

In Fig.~\ref{mwg} we represent the O/H radial gradient as a function of redshift for the six models for MWG calculated 
in  Moll{\'a} et al. (2017) with the different prescriptions of the HI to H$_{2}$ conversion as explained. In top panels we show these gradients as dex\,kpc$^{-1}$,
computed in panel a) with the whole radial range for which we have calculated the models. In panel b) we have the gradients calculated
with regions for which $R\le\,2.5\,R_{\rm eff}$. We may see that in this last case the gradient is basically constant along $z$ with 
very small differences among models. In both panels we have included the MWG data which give this evolution along $z$, PN from 
Stanghellini \& Haywood (2010) and Maciel \& Costa (2013); open clusters from Cunha et al. (2016); 
and stellar abundance from Anders et al. (2016). The present time values are from Henry et al. (2010), Rupke et al. (2010), Genovali et al. (2014) and
the one compiled by Moll{\'a} et al. (2015), as labelled. We have also shown the cosmological simulation result for a Milly Way-like galaxy from Gibson et al. (2013, G13). We see that, in order to reproduce the data, it is necessary to compute the gradient within the optical radius. In bottom panels we show the gradients obtained using a normalized radius ($R/R_{\rm eff}$).  In panel c), where we use
again all radial range, we see a different behavior between STD and MOD models (which use an efficiency to form molecular clouds) and
all the others using a prescription to convert HI in H$_{2}$ which depend on the gas, stars or total density or/and on the dust through the
metallicity.  In these last cases the effective radius increases more slowly than in our standard models, thus producing a strong radial gradient when
regions out of the optical disk are included for the fit. In panel d) , where only regions with  $R\le\,2.5\,R_{\rm eff}$ are used, we find again a very
constant  radial gradient along the redshift. In fact, this value is in very good agreement with the one found by S{\' a}nchez et al. (2014, S14) as a common 
gradient for all the CALIFA survey galaxies.

\begin{figure} 
\begin{center}
\includegraphics[width=9.0cm,,angle=-90]{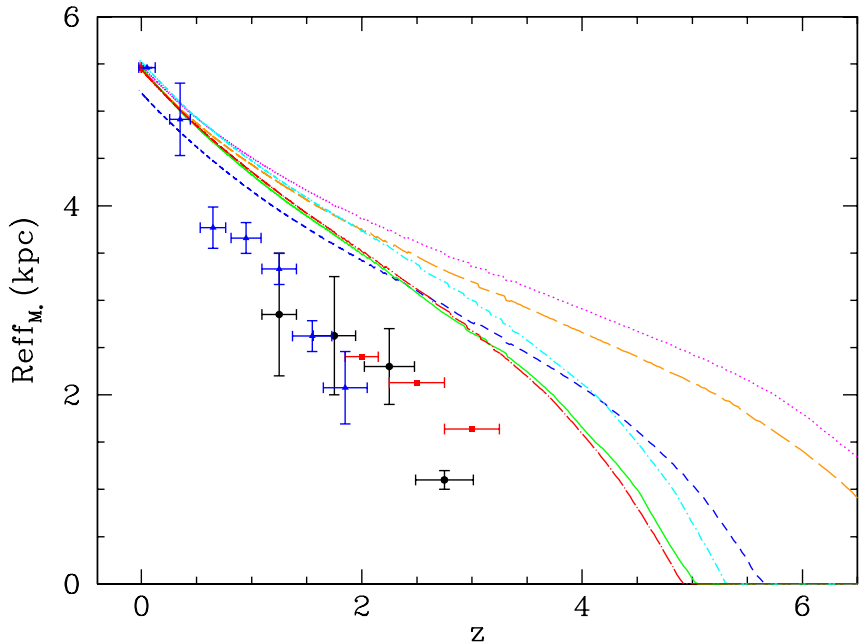}
\caption{Evolution of the effective radius R$_{\rm eff}$ with redshift $z$. Data are from  Trujillo et al (2007), Buitrago et al. (2008),  and  Margolet-Bentatol  et al. (2016),
as blue triangles, red squares, and black dots, respectively.}
\label{radio}
\end{center}
\end{figure}

The grow of the disk in the different models is shown in Fig.~\ref{radio} with data as labelled. We see, as said before, that ASC model is the one where the radius
increases more slowly while the STD and MOD models started very early to show a large disk. GNE, BLI and KRU show an intermediate behavior. Although the data we show in Fig.~\ref{radio} refer mainly to bulges and disks from early-type galaxies, it seems clear that ASC is the model closest to the observations. 

\subsection{Evolution of the O/H radial gradient in spiral and low mass galaxies}

\begin{figure}  %%%%%%%%%%%%%%%%%%%%%%FIGURE 2ab %%%%%%%%%%%%%%%%%%%%%%%
\begin{center}
\includegraphics[width=9.0cm,angle=-90]{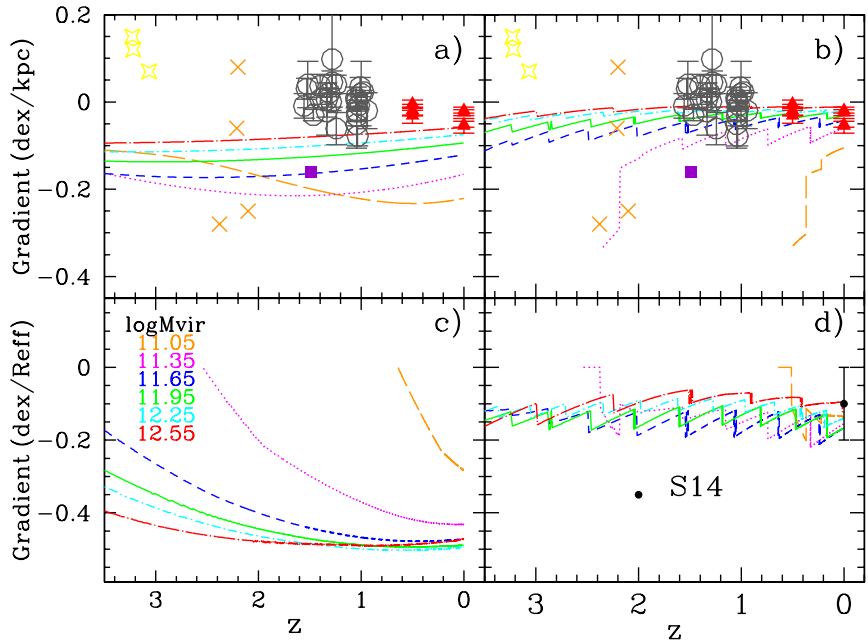}
\caption{ Evolution of the O/H radial gradient with redshift $z$ for several galaxy models with different virial masses ($M_{\rm vir}$), measured as: dex\,kpc$^{-1}$ (top panels) and dex\,R$_{\rm eff}^{-1}$ (bottom panels). Left panels show the gradients obtained using the whole radial range available in the simulations. Right panels show gradients computed only using regions within the optical disk defined as $\rm R\le 2.5\,R_{\rm eff}$. 
Data correspond to the observations by Cresci et al. (2010, yellow stars), Yuan et al. (2011, purple square),  Queyrel et al. (2012 grey open dots), 
 Jones et al. 2013 (orange crosses), and   Magrini et al. (2016, red points with error bars).}
\label{sps}
\end{center}
\end{figure}

In Fig.~\ref{sps} we show the radial gradients computed for different galaxy masses, as labelled, in a similar way than Fig.~\ref{mwg}. In panel a), as in Fig.1a, the radial gradient computed for all radial regions is shown. It is clear that each galaxy has its own evolution being the smallest one which shows the most different behavior.
Each galaxy has a different radial gradient, with the most massive ones showing the flattest distributions ($\sim -0.05\,{\rm dex\,kpc}^{-1}$ for all $z$), while the smallest has the steepest gradient  ($\sim -0.20\,{\rm dex\, kpc}^{-1}$). However, when only radial regions within the optical radius are used to compute the radial gradient a very different behavior arises: all gradients are approximately constant with $z$ for galaxies with $\log(M_{\rm vir})\geq 11.65$, although with the same behavior than before: the more massive the galaxy, the flatter the gradient. In the lowest masses galaxies, it is evident the moment in which the disk begins to grow: at $z=2.5$ for $\log{M_{\rm vir}}=11.35$ and at $z=0.5$ for $\log{M_{\rm vir}}=11.05$.  

When we represent the gradients measured as function of the effective radius, we see that they steepen with decreasing $z$ when all radial regions are used (Fig. 3c) and, again, a very smooth evolution along $z$ for all galaxies appears when only the optical disk is used to fit the gradient (panel d).
The average value in this case is $\sim -0.13\,{\rm dex\,kpc}^{-1}$, as the value found by S{\'a}nchez et al. (2014), supporting their claim that a universal radial gradient appears for all galaxies. 

\begin{figure} 
\begin{center}
\includegraphics[width=7cm,,angle=-90]{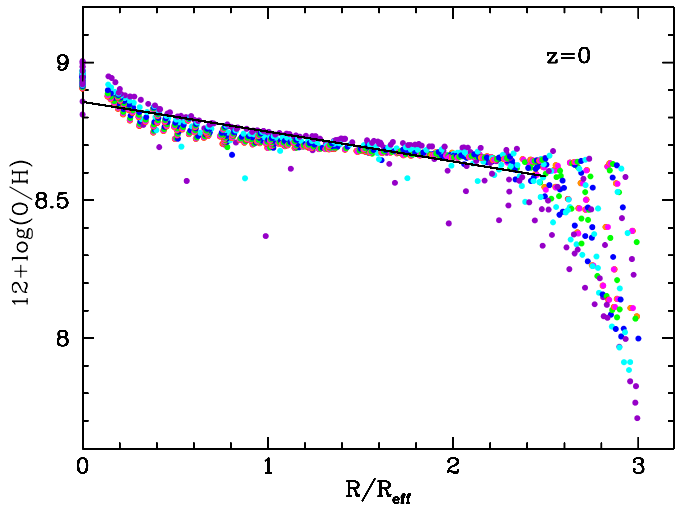}
\caption{The O/H radial distribution for the present time obtained with our series of models  for all galaxy masses and six different efficiencies to form stars with values $\epsilon_{*} \ge 0.002 $,  as a function of the normalized radius R/R$_{\rm eff}$. Each color shows a different efficiency $\epsilon_{*}$.} 
\label{grad-reff}
\end{center}
\end{figure}

A common radial gradient is easily obtained drawing O/H for the present time as a function of R/R$_{\rm eff}$ for all galaxy masses and efficiencies (larger than 0.002) in a same plot, as we show in Fig.\ref{grad-reff}. We see that effectively, such as S{\'a}nchez et al. (2014) found, a same radial gradient is obtained for
all models, when R/R$_{\rm eff} \le 2.5$, with a dispersion given by the differences in the star formation efficiencies around an average radial distribution.

\begin{figure} 
\begin{center}
\includegraphics[width=7cm,,angle=-90]{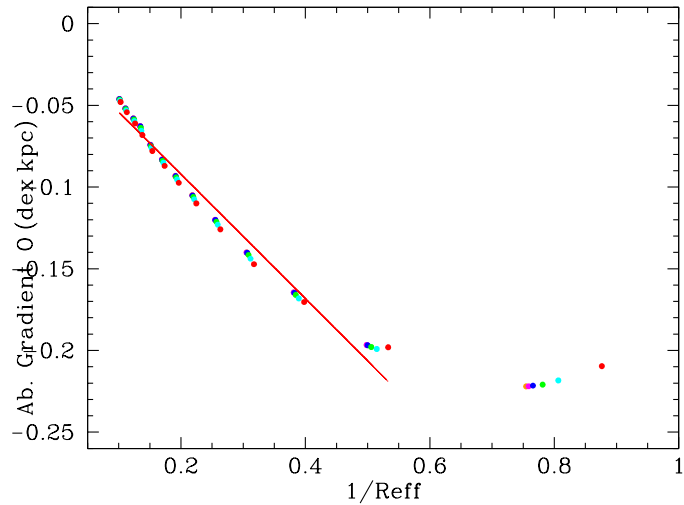}
\caption{The O/H radial gradient measured as dex\,kpc$^{-1}$,  as a function of the inverse of the effective radius, 1/R$_{\rm eff}$. Each color shows a different 
efficiency $\epsilon_{*}$.} 
\label{corr}
\end{center}
\end{figure}

Since it seems quite evident that the radial gradient is a scale effect due to the star formation rate which is measuring the stellar disk growth, we would expect
a correlation between this O/H radial gradient measured as dex\,kpc$^Ð{1}$ and the scale length of the disk or any other quantity defining the size of the disk.
We plot in Fig.~\ref{corr}, right panel, this correlation for all our models with different galaxy masses and with six different values for the efficiencies to form stars
from molecular clouds, which are coded with different colored dots. The correlation is clear for all effective radii larger than 1.25\,~kpc. If the effective radius is
smaller than this value, our code, working with radial regions of 1\,kpc wide, is not able to calculate a radial gradient nor an effective radius.
This theoretical correlation supports the observational  one found by Bresolin \& Kennicutt (2015)  with the radial gradient and the scale length of the disks
(their  Fig.~3).  These authors claim in that work that all galaxies, even the low surface brightness galaxies, share a common abundance radial gradient when this one
is expressed in terms of the exponential disk scale-length (or any other normalization quantity).

\section{Conclusions}

The conclusions can be summarized as:

\begin{itemize}
\item A grid of chemical evolution models with 76 different total dynamical masses in the range 10$^{10}$ to 10$^{13}$ M$_{\odot}$ is calculated.
\item 10 values of efficiencies $\epsilon_{*}$ to form stars from molecular clouds are used with values $ 0 < \epsilon_{*} < 1$. But we find that useful values
are only the first six-seven of them with $\epsilon_{*} > 0.002$.
\item The best combination IMF from Kroupa et al (2002) $+$ Gavilan et al. (2006) $+$ Chiefi \& Limongi (2003,2004) yields, is used.
The stellar yields $+$ IMF may modify the absolute abundances on a disk, but they do not change the radial slope of the abundance distributions of disks.
\item Using Shankar et al. (2006) prescriptions for $M_{\rm halo}/M_{\rm disk}$, we obtain the necessary infall rates to reproduce the radial profiles of  galaxy disks
\item Different prescriptions for the conversion of HI to H$_{2}$  are used finding that the ASC model is the best one.
\item The slope of the oxygen abundance radial gradient for a MWG-like model when it is measured for $R<2.5\,R_{\rm eff}$ has a value ~$-0.06$\,dex\,kpc$^{-1}$, which is around $-0.10$\, dex\,R$_{\rm eff}^{-1}$ when it is measured using a normalized radius.
\item This same slope is  also obtained for all efficiencies and all galaxy masses  in excellent agreement with CALIFA results, supporting the idea of a universal radial gradient for all galaxies when measured as a function of a normalized radius.
\item The slope do not changes very much along $z$ when the infall rate is as smooth as we have obtained recently, compared with old models with a stronger evolution.
\end{itemize}

\label{discussion}

\acknowledgments 
This work has been supported by DGICYT grant AYA2013-47742-C4-4-P. This work has been sup- ported financially by grant 2012/22236-3 from the S{\~a}o Paulo Research Foundation (FAPESP). This work has made use of the computing facilities of the Laboratory of Astroinformatics (IAG/USP, NAT/Unicsul), whose purchase was made possible by the Brazilian agency FAPESP (grant 2009/54006- 4) and the INCT-A. MM thanks the kind hospitality and wonderful welcome of the Jeremiah Horrocks Institute at the University of Central Lancashire, the E.A. Milne Centre for Astrophysics at the University of Hull, and the Instituto de Astronomia, Geof{\'\i}sica e 
Ci{\^e}ncias Atmosf{\'e}ricas in S{\~a}o Paulo (Brazil), where this work was partially done.


\begin{references}
\reference Anders, F., Chiappini, C., Minchev, I., et al.\ 2016,  A\&A, submitted (arXiv:1608.04951) 

\reference Blitz, L., \& Rosolowsky, E.\ 2006, ApJ, 650, 933 

\reference Bresolin F., Kennicutt R.~C., 2015, MNRAS, 454, 3664 

\reference Buitrago F., Trujillo I., Conselice C.~J., Bouwens R.~J., et al. , 2008, ApJ, 687, L61 

\reference Chieffi A., Limongi M., 2004, \textit{ApJ}, 608, 405

\reference Cresci, G., Mannucci, F., Maiolino, R., et al.\ 2010, \textit{Nature}, 467, 811

\reference Costa, R.~D.~D., Cavichia, O., Maciel, W.~J., 2013, \textit{IAUS}, 289, 375 

\reference Cunha, K., Frinchaboy, P.~M., Souto, D., et al.\ 2016, \textit{Astronomische Nachrichten}, 337, 922

\reference Ferrini, F., Molla, M., Pardi, M.~C., Diaz, A.~I., 1994, \textit{ApJ}, 427, 745

\reference Gavil{\'a}n, M., Buell, J.~F., \& Moll{\'a}, M., 2005, \textit{A\&A}, 432, 861

\reference Gavil{\'a}n, M., Moll{\'a}, M., \& Buell, J.~F., 2006, \textit{A\&A}, 450, 509

\reference Gibson, B.~K., Pilkington, K., Brook, C.~B., Stinson, G.~S., \& Bailin, J.\ 2013, \textit{A\&A}, 554, A47

\reference Gnedin N.~Y., Kravtsov A.~V., 2011, ApJ, 728, 88 

\reference Henry, R.~B.~C., \& Worthey, G.\ 1999, \textit{PASP}, 111, 919

\reference Henry, R.~B.~C., Kwitter, K.~B., Jaskot, A.~E., et al.\ 2010, \textit{ApJ}, 724, 748

\reference Iwamoto, K., Brachwitz, F., Nomoto, K., et al.\ 1999, \textit{ApJS}, 125, 439

\reference Jones, T., Ellis, R.~S., Richard, J., \& Jullo, E.\ 2013, \textit{ApJ}, 765, 48 

\reference Kroupa, P., 2002, \textit{Sci}, 295, 82

\reference Krumhold, M.R., McKee, C.F., \& Tumlinson, J., 2008, ApJ, 689, 865

\reference Krumhold, M.~R., McKee, C.~F., \& Tumlinson, J., 2009, ApJ, 693, 216
 
\reference Limongi, M., \& Chieffi, A., 2003, \textit{ApJ}, 592, 404

\reference Maciel, W.~J., Costa, R.~D.~D., \& Uchida, M.~M.~M., 2003, \textit{A\&A}, 397, 667 

\reference Maciel, W.~J., \& Costa, R.~D.~D.\ 2013, \textit{RMxA\&Ap}, 49, 333

\reference Magrini, L., Coccato, L., Stanghellini, L., Casasola, V., Galli, D.\, 2016, \textit{A\&A}, 588, A91 

\reference Margalef-Bentabol B., Conselice C.~J., Mortlock A., Hartley W.,  et al., 2016, MNRAS, 461, 2728 

\reference Moll{\'a}, M., Ferrini, F., \& Diaz, A.~I.\ 1996, \textit{ApJ}, 466, 668

\reference Moll{\'a}, M., Ferrini, F., \& Diaz, A.~I.\ 1997, \textit{ApJ}, 475, 519

\reference Moll{\'a}, M. \& D{\'{\i}}az, A.I.\ 2005, \textit{MNRAS}, 358, 521 (MD05)

\reference Moll{\'a}, M., Cavichia O., Gavil{\'a}n M., \& Gibson B.~K.\ 2015, \textit{MNRAS}, 451, 3693

\reference Moll{\'a}, M., D\'{\i}az, A.~I., Gibson, B.~K.,  et al., 2016, \textit{MNRAS}, 462, 1329

\reference Moll{\'a}, M., Ascasibar, Y., D\'{\i}az, A.~I., \& Gibson, B.~K., 2017, \textit{MNRAS}, to be submitted

\reference Pilkington, K., Few, C.~G., Gibson, B.~K., et al.\ 2012, \textit{A\&A}, 540, A56

\reference Queyrel, J., Contini, T., Kissler-Patig, M., et al.\ 2012, \textit{A\&A}, 539, A93

\reference Rupke, D.~S.~N., Kewley, L.~J., \& Chien, L.-H.\ 2010, ApJ, 723, 1255

\reference Salucci, P., Lapi, A., Tonini, C., Gentile, G., et al.., 2007, \textit{MNRAS}, 378, 41

\reference S{\'a}nchez, S.~F., Rosales-Ortega, F.~F., Iglesias-P{\'a}ramo, J., et al.\ 2014, \textit{A\&A}, 563, A49

\reference Shaver, P.~A., McGee, R.~X., Newton, L.~M., Danks, A.~C.,  et al.\ 1983, \textit{MNRAS}, 204, 53

\reference Stanghellini, L., \& Haywood, M.\ 2010, \textit{ApJ}, 714, 1096 

\reference Trujillo I., Conselice C.~J., Bundy K., Cooper M.~C.,  et al., 2007, MNRAS, 382, 109 

\reference Xiang, M.-S., Liu, X.-W., Yuan, H.-B., et al.\ 2015, \textit{Research in Astronomy and Astrophysics}, 15, 1209 

\reference Yuan, T.-T., Kewley, L.~J., Swinbank, A.~M., et al. .\ 2011, \textit{ApJL}, 732, L14

\reference Zaritsky, D., Kennicutt, R.~C., Jr., \& Huchra, J.~P.\ 1994, \textit{ApJ}, 420, 87

\end{references}
\end{document}